\documentclass[useAMS,usenatbib]{mn2e}
\pdfoutput=1
\usepackage[english]{babel}
\usepackage{amssymb}
\usepackage{enumerate}
\usepackage{epsfig,amsmath}
\usepackage{graphics}
\usepackage{boldtensors}

\DeclareMathSymbol{\Natural}{\mathbin}{AMSb}{"4E}
\DeclareMathSymbol{\Integer}{\mathbin}{AMSb}{"5A}
\DeclareMathSymbol{\Real}{\mathbin}{AMSb}{"52}
\DeclareMathSymbol{\Rational}{\mathbin}{AMSb}{"51}
\DeclareMathSymbol{\Imaginary}{\mathbin}{AMSb}{"49}
\DeclareMathSymbol{\Complex}{\mathbin}{AMSb}{"43}

\def\Pr{{\rm Pr}}

\title[Extreme-value methods for periodograms]{Extreme-value modelling for the significance assessment of periodogram peaks}
\author[M. S\"{u}veges]{M. S\"{u}veges$^{1}$\thanks{E-mail: Maria.Suveges@unige.ch} \\
$^{1}$ISDC Data Centre for Astrophysics,  Department of Astronomy, University of Geneva, Chemin d'Ecogia 16, \\ $\quad$ CH-1290 Versoix, Switzerland; \\ $\;$Chair of Statistics, Station 8, Ecole Polytechnique F\'ed\'erale de Lausanne, CH-1015 Lausanne, Switzerland}
\begin{document}

\date{}

\pagerange{\pageref{firstpage}--\pageref{lastpage}} \pubyear{2012}

\maketitle

\label{firstpage}

\begin{abstract}
I propose a new procedure to estimate the False Alarm Probability, the measure of significance for peaks of periodograms. The key element of the new procedure is the use of generalized extreme-value distributions, the limiting distribution for maxima of variables from most continuous distributions. This technique allows reliable extrapolation to the very high probability levels required by multiple hypothesis testing, and enables the derivation of confidence intervals of the estimated levels. The estimates are stable against deviations from distributional assumptions, which are otherwise usually made either about the observations themselves or about the theoretical univariate distribution of the periodogram. The quality and the performance of the procedure is demonstrated on simulations and on two multimode variable stars from Sloan Digital Sky Survey Stripe 82.
\end{abstract}


\section{Introduction} \label{sec_intro}

Several research fields of astronomy rely crucially on the analysis of periodic phenomena. For instance, asteroseismology depends on a reliable identification of excited oscillation modes in variable stars, in order to match theoretical stellar models to observed data (\citealt{smolecmoskalik07,grigahceneetal10, balonadziembowski11, antocietal11}, to mention only a few). As another example, extrasolar planetary systems are found by periodic modulations in the light or the radial velocity of their central stars \citep{cummingetal99, udryetal07, mayoretal09, dawsonfabrycky10}. Thus, the detection and analysis of periodic signals in astronomical time series receives much attention in the literature (for a concise summary of the theoretical background, see \citealt{schwarzenberg-czerny98}).

The general principles of the detection are those of statistical hypothesis testing and model selection. A null hypothesis $H_0$ of no periodic signal in the observed time series is tested against the alternative of the presence of a periodic deterministic component.  $H_0$ supposes most often that the observations are white noise (eventually Gaussian) around a constant mean, though the absence of periodicities do not exclude eventually other error distributions, errors with nontrivial time series structure and non-periodic stochastic processes like random walks. The white noise null hypothesis is made plausible by pre-processing the data:  pre-selection of candidates (e.g. quasar/star separation in a deep survey to filter out cases where random walks may be a good model instead of periodic light variations), and a careful assessment whether independence or uncorrelatedness is a sufficiently good approximation for the errors. The alternative hypothesis is formalised by a series of more complex models $\mathcal{M}_f$ indexed by the frequency $f$, consisting of a periodic signal at $f$ and noise. Both the null model  $\mathcal{M}_0$ and the collection $\mathcal{M}_f$ are fitted to the data, and a test statistic $\theta(f)$ is computed for all models, which quantifies the improvement yielded by $\mathcal{M}_f$ over $\mathcal{M}_0$. This collection of the test statistics as a function of $f$ is called the periodogram. The frequency at which the largest improvement is achieved is accepted as the most likely frequency of an eventual periodic component. Then the decision, whether the object shows a periodic oscillation or not, is based on the significance assessment of the model improvement: the probability is computed that the realized periodogram maximum or a higher value is produced under the null hypothesis. This probability is termed the False Alarm Probability (FAP).

 This assessment is a multiple testing situation: as we do not know the frequency of the sought oscillation in advance, we compute the test statistic often at hundreds of thousands of frequencies. Thus, the single-value distribution $F$ (the marginal distribution) of the individual test statistics is not directly applicable to compute the FAP. Instead, we must find the distribution $G$ of the maximum of a large set of test statistics.

The idea which is most commonly used in astronomy to obtain $G$ and the FAP is based on elementary probability calculations for the maxima of $M$ independent variables with common distribution function $F$, yielding the formula $G(z) = F(z)^M$ \citep{scargle82, hornebaliunas86, schwarzenberg-czerny98}. 
This formula has the great merits of  being simple, quick to compute and easily applicable even for large surveys in an automated way.

However, there are numerous issues with it, both on the theoretical side and in practice. Astronomical time series are typically irregularly spaced, sparsely sampled in time domain, and the periodogram is computed on an oversampled frequency grid. As a result, there is no set of independent or uncorrelated frequencies in the astronomical periodogram, $M$ loses its direct meaning, and must be interpreted as an effective number of hypothetical independent frequencies. Thus, the formula itself becomes an ad hoc statistical model rather than a well-founded approximation. On the practical side, the form $F(z)^M$ is highly unstable, sensitive to misestimation of both of its components, and is not adapted for statistical inference. A more reliable FAP estimate, with sound statistical foundations and with a possibility to assess the uncertainty in the FAP level estimate would be desirable.

This paper proposes to avoid the shortcomings of the formula $F(z)^M$ by the application of extreme-value statistics. Statistical theory proved that maxima of random variables follow a simple  3-parameter limiting distribution, called the generalized extreme-value family (GEV; \citealt{fishertippett28, gnedenko43}). The validity range of this limit theorem is very broad:  it encompasses practically all continuous distributions and dependent variables too, under some mild conditions \citep{leadbetter74}. Regardless of the underlying marginal distribution of the variables, it has one common parametric form, estimable with standard methods, and provides a stable, mathematically well-founded model for extrapolation to the levels required by the FAP.  Instead of the unstable formula $F(z)^M$, the use of the GEV is standard practice in most sciences, industries, finance or economy that are concerned with risk estimation of rare events like hurricanes, floods, droughts, internet traffic failures, or market crashes (for a range of applications, see e.g. \citealt{finkenstadtrootzen}). A first use of extreme-value theory for periodograms in astronomy was proposed by \citet{baluev08}, providing an upper bound on FAP under some strict conditions on the distribution of the periodogram and under a low level of aliasing and spectral leakage. 

The procedure proposed here fits a GEV model to the tail of the distribution of the periodogram under the null hypothesis of the time series, that is, it is white noise. Then this is used to find either the FAP of an observed periodogram peak, or critical levels corresponding to pre-specified FAP levels. First, we construct a large number of noise time series corresponding to the null hypothesis. In the next step, we compute part of the periodograms of the constructed time series, and find the maxima of these partial periodograms. Finally, the parameters of the GEV distribution of these maxima are estimated, and the obtained GEV is used to extrapolate to the desired high levels. 

In Section \ref{sec_theory}, we first give a brief summary about frequency analysis and its particularities in astronomy, and discuss in detail the consequent problems in the statistical hypothesis testing for the significance of periodogram maxima. Then we present the basics of extreme-value theory and inference, and discuss its potential and difficulties. Section \ref{sec_proc} describes the proposed procedure, and explains the arguments motivating the applied techniques. Section \ref{sec_res} shows the performance of the procedure on simulations, while Section \ref{sec_rr} applies the methods to two variable stars from Stripe 82 of the Sloan Digital Sky Survey with an RR Lyrae-like primary frequency. Section \ref{sec_disc} gives a summary of the results.

\section{Statistical background}\label{sec_theory}
 
\subsection{Frequency analysis}\label{subsec_fou}

Suppose we have an observed time series $X_1, \ldots, X_N$ with $N$ points, e.g. magnitudes or radial velocities of a star, measured at times $t_1, \ldots, t_N$. The exposure time and other practical factors limit the precision of the epochs, and there can be found a greatest common divisor $\delta t$ of the time differences $t_i - t_j, \; i,j = 1, \ldots, N$, such that we can regard the measurements as taken on a dense time grid of resolution  $\delta t$, consisting of epochs $0, 1\delta t, 2\delta t, \ldots, T\delta t$ \citep{eyerbartholdi99}. The observed sequence can then be regarded as a time series taken on a regular grid, with a scarce minority of known observations at $t_1, \ldots, t_N$ and with an overwhelming majority of missing values at other epochs of the grid. In the sequel, the term 'irregularly or unevenly sampled time series' will refer to such an observational sequence. 
 
Periodic signals in an evenly sampled time series $X_1, \ldots, X_T$ are usually detected by the means of the periodogram, an estimate of the spectral density of the time series, on a fixed frequency grid $\mathcal{F}$ between 0 and  $f_{\max}$, where $f_{\max} \leq f_{\rm Nyquist} = 1/2\delta t$ \citep{eyerbartholdi99}. The classical periodogram is defined as
\begin{eqnarray} \label{eq:pgram}
I_{T,X}(f) = \frac{1}{T\delta t} \left| \sum_{j = 1}^T e^{-i2\pi f j\delta t} X_j \right|^2.
\end{eqnarray}

For evenly observed signals, the fundamental method to estimate the periodogram at the Fourier frequencies is the fast Fourier transform. For the irregularly sampled case, many methods can be found in the astronomical literature: the Deeming method \citep{deeming75} that can be regarded as the direct generalization of \eqref{eq:pgram} to arbitrary times; PDM-Jurkevich \citep{jurkevich71,stellingwerf78,dupuyhoffman85}; string length \citep{clarke02}; SuperSmoother \citep{friedman84, reimann94}; Keplerian periodograms \citep{cumming04};  Lomb-Scargle and its extension, the generalized Lomb-Scargle method (\citealt{lomb76,scargle82, zechmeisterkurster09}); the FastChi2 of \citet{palmer09}; and for photon arrival time series, methods based on Rayleigh's and Kuiper's tests \citep{paltani04}. The extremum (most often the maximum) of the estimated periodogram indicates the most likely frequency of the object, and statistical hypothesis testing is used to decide whether the periodic component is significant or not.

This  hinges on the knowledge of the distribution $G$ of the maximum. Most arguments for its derivation rely directly or indirectly on the relationship $F(z)^M$, which follows from independency assumptions \citep{scargle82, hornebaliunas86}. If there is a set of independent random variables $Z_1, \ldots, Z_M$ with common distribution function $F$, then the distribution of their maximum can be derived as
\begin{eqnarray} \label{eq:FadM}
 &\Pr & \!\!\!\!\!\! \left( \max \{Z_1, \ldots, Z_M\} \leq z  \right) \nonumber  \\
 &      &= \Pr  \left( Z_1 \leq z, \ldots, Z_M \leq z  \right) \\
  &     &= \Pr  \left( Z_1 \leq z  \right) \times \ldots \times \Pr  \left( Z_M \leq z  \right) = F(z)^M, \nonumber 
\end{eqnarray}
 where the second equality is true only if the variables  $Z_1, \ldots, Z_M$ are independent; otherwise, the joint probability cannot be decomposed into a product of the marginal probabilities. 
 
 So far, few easy-to-use alternatives were proposed to this formula, and most procedures for FAP estimation rely on it (though there are bootstrap-based alternatives as in \citealt{paltani04} and \citealt{schwarzenberg-czerny12}, or procedures using empirically derived reference distributions as in \citealt{koeneyer02}). However, there are several drawbacks. 

\begin{itemize}

\item \textit{No independent set of frequencies exists for an irregular sparse time sampling}. The irregular sampling introduces non-vanishing correlations between sine functions of different frequencies, so the system of mutually orthogonal frequencies disappears. This entails the loss of any independent frequency systems even in a Gaussian case. Thus, an appropriate derivation of the distribution of the maximum should use the joint multivariate distribution of the periodogram.

\item \textit{This joint multivariate distribution is degenerate.} We use $N$ observations to compute $n \gg N$ test statistic values $\theta(f_i) \; (i = 1, \ldots, n)$. Any mapping $h: \Real^N \longrightarrow \Real^n$, defined on the space of the observed random variables $X_1, \ldots, X_N$ to compute $n$ values $Z_1, \ldots, Z_n$, produces degenerate joint probability distributions if $n > N$ ($N/2$ for periodograms, as we lose the phase information during the computation). The testing situation is therefore not just mildly, but radically different from the basic assumptions of \eqref{eq:FadM}. 

\item \textit{The marginal distribution $F(z)$ is usually unknown.} 
Theoretically derived approximate marginal distributions for the periodogram are usually based on the asymptotic theory of least squares estimation or the Central Limit Theorem, using the orthogonality of the basis functions. However, in many cases orthogonality does not hold. An example is the inclusion of a constant beside $\sin 2\pi f t$ and  $\cos 2\pi f t$ in the generalized Lomb-Scargle method. The latter two functions, providing the basis for Fourier analysis and the (non-generalized, zero-mean) least squares method, can be made orthogonal for irregular sampling by a phase shift \citep{scargle82}. Depending on normalization and the test statistic used, the marginal distribution of the periodogram can then be shown to be either the $\chi^2$ or the Fisher-Snedecor or the beta distribution (the first is for known spectrum normalization, the second two are for empirically estimated noise levels depending on test statistic; see \citealt{schwarzenberg-czerny98}). The generalized Lomb-Scargle  method adds the constant function $f(t) = 1$ to these two basis functions, allowing for the presence of an unrestricted floating mean at all frequencies. This extended basis is not orthogonal. Orthogonalization would require a time-consuming Gram-Schmidt procedure \citep{ferraz-mello81} that is rarely done in practice. Consequently, the $\chi^2$, the  Fisher-Snedecor or the beta distribution can be accepted only as approximate models for the periodogram distribution. 

\begin{figure}
\begin{center}
\includegraphics[scale=.58]{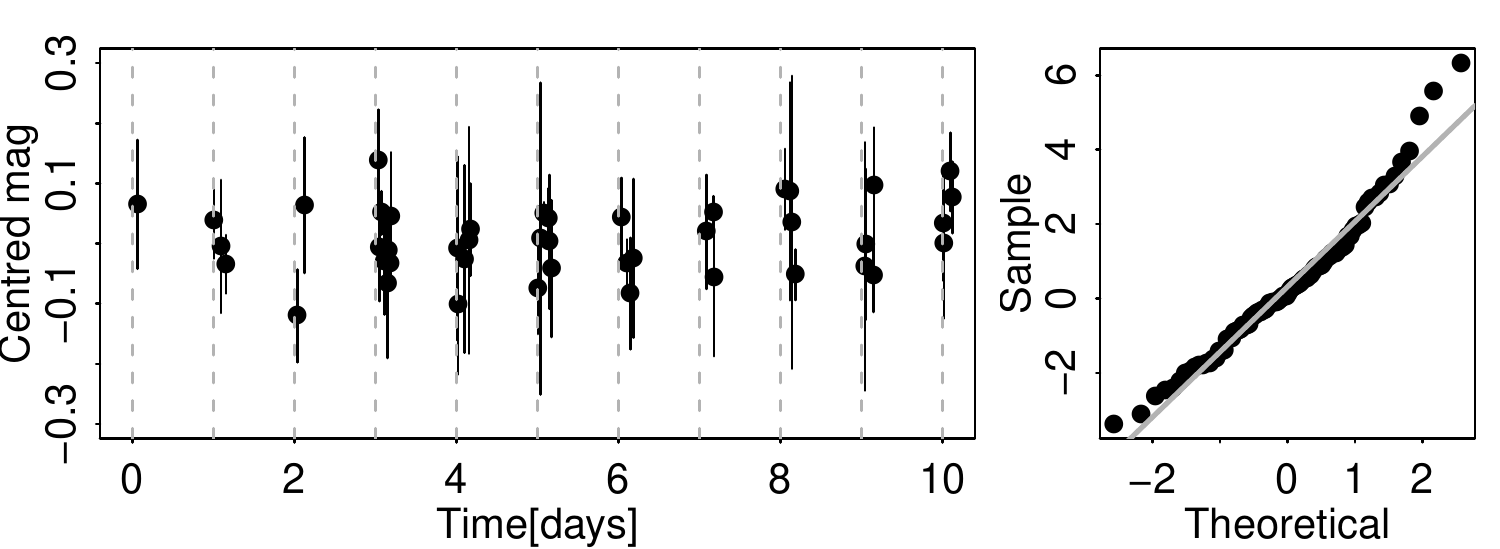}
\caption{A simulated time series (sinusoidal signal and independent Gaussian errors with SNR $= 0.5$ and frequency $3.379865 \; d^{-1}$) and its quantile-quantile plot. The signal is undetectable in this case.}
\label{fig:nonnormality}       
\end{center}
\end{figure}

\item \textit{Theoretical approximations may be anyway rough in some cases:}  if there are only a small number of observations and strong departures from normality in their tail. Approximate marginal distributions for the periodogram in such cases can be poor, as they are only asymptotically valid when the observations themselves are not Gaussian. Such a case is illustrated in Figure \ref{fig:nonnormality}: the standardized Gaussian quantile-quantile plot (right panel) of the observed sequence (shown on the left) shows strong deviations from the straight line representing a true Gaussian distribution, especially in the high end.  

\item \textit{It is necessary to estimate $M$ and $F(z)$}. $M$ does not correspond to the count of any interpretable or identifiable variable in a simplified model that could be straightforwardly derived from the testing situation, so its estimation is usually based on fits to simulations including strong assumptions. The estimated value of $M$ is most often larger than the number of observations \citep{hornebaliunas86, frescuraetal08, schwarzenberg-czerny12}, which is impossible in the context of an independent model. This hints at the ad hoc nature of the approximation $F(z)^M$: the absence of independence and the degeneracy in the periodogram makes the simplification to an equivalent independent system hard to imagine. 

\item \textit{The formula  $F(z)^M$ is very sensitive to tiny changes in both $M$ and $F$.} Extremely high quantiles of $F$ (of the order of $F(z) = 0.9999$) must be precisely estimated in order to obtain even moderate FAP levels: even with $M = 25$ (an unusually low value) and FAP$=0.01$, we need $z$ such that $F(z) = 0.9996$, since FAP $=1-F(z)^M = 1-0.9996^{25}$. Neither theoretical nor estimated marginal distribution functions $F$ are reliable in this range. Theoretical approximations are usually tailored to the center of the distribution, while estimations practically never have data at such extreme levels, so extrapolation conveys information about our preliminary assumptions rather than about the true distribution at the high end. The form $F(z)^M$, with $M$ in the exponent of a highly uncertain tail distribution, can easily result in an erroneous estimation. 

\item \textit{Moreover, the distribution $F(z)^M$ itself becomes degenerate when $M$ increases}. This causes a situation contradicting a fundamental paradigm in statistical inference, namely, that the distribution of the test statistic should tend to a stable well-defined distribution when the number of observations $N$ increase, and in general, the quality of the estimates should improve. However, in the analysis of astronomical periodograms the increase of $N$ usually causes also $M$ to increase, and this creates a catastrophic situation for inference, for the following reason. Let $z_+$ denote the endpoint of the marginal distribution $F(z)$ of the periodogram. This endpoint can be finite for a finite-domain distribution such as the uniform or the beta distribution, or $\infty$ if it has no finite upper limit, such as the Fisher-Snedecor or the $\chi^2$ distribution. 
Since $F(z) < 1$ for all $z < z_+$, and  $F(z) = 1$ only for  $z = z_+$, $F(z)^{M}$ tends to 0 for all $z < z_+$, and to 1 only for  $z = z_+$. Thus, when $z_+$ is finite, $F(z)^{M}$ tends to a step function with an extremely steep rising part just before $z_+$, and to give a good probability estimate, we need a precision in $z$ tending to infinity. When $z_+$ is infinite, $F(z)^{M} \rightarrow 0$ everywhere on the real line, and $z$ values corresponding to the necessary near-one probability levels run out to infinity. This instability of $ F(z)^{M}$ implies that no stable limit distribution can be given for the estimated FAP levels, and no regular statistical inference can be obtained for the quantile estimates. Stabilization is necessary, and this is what is obtained by extreme-value theory.
\end{itemize}

\subsection{Generalized extreme-value distributions}\label{subsec_evd}

Extreme-value theory deals with the statistical analysis of low-probability events. Its development has been motivated by the need of estimating the probability and the level of rare, but possibly catastrophic events, such as the hurricane Katrina in 2005; the 3-day rainstorm on December 14-16, 1999, which caused around 30,000 deaths in the coastal areas of Venezuela;  or financial market crashes that can have serious impact on economic stability. The theory has been summarized in several books (e.g. \citealt{leadbetter, resnick87, embrechtsetal, colesbook, statofext, dehaanferreira}).  Its cornerstone theorem yields a family of asymptotically valid limiting distributions for the (normalized) maxima $Z_{\mathrm{max},n}$ of a large number of random variables $Z_1, \ldots, Z_n$, identically distributed according to a continuous distribution $F(z)$, when $n \longrightarrow \infty$ (the normalization constants are irrelevant for estimation, as they are merged into the parameters of the distribution). The theorem is valid for the maxima of not only independent, but dependent variables too, if the dependence decays between increasingly separated extremely large variables when $n \longrightarrow \infty$. The extremal types theorem \citep{fishertippett28,gnedenko43} states the form of this limiting family, called the generalized extreme-value distribution (GEV):
\begin{eqnarray} \label{eq:gev}
G(z) &= & {\rm Pr} \{ Z_{\max,n} \leq z  \}  \nonumber \\
         &= &\exp  \left\{ - \left( 1 + \xi \frac{z - \mu}{\sigma} \right)^{-1/\xi} \right\}, \\
  &   & \xi \in \Real, \quad \mu \in \Real, \quad \sigma > 0,  \nonumber
\end{eqnarray}
where $z$ is such that $1+\xi(z-\mu)/\sigma > 0$. The exact formulation of the extremal types theorem can be found in the references given above.

The GEV family constitutes the limiting family for the maxima of nearly all continuous distributions. 
The parameter $\xi$, called shape, is related to the tail decay of the underlying distribution of $F(z)$, and divides the GEV family into three well-separated subfamilies. Negative shape parameters provide distributions of maxima of variables with a finite upper boundary, for example the uniform or the beta distribution. In this case, $z < \mu - \xi \sigma$, and there is zero probability to obtain maximum values higher than this limit. With a positive shape parameter, there is no upper limit, the probabilities in the right tail of the GEV  decay slowly as a power law, and there is considerable chance for the maximum to reach very large values. This is the limit distribution of maxima from heavy-tailed laws like the Student's $t$, or the Cauchy distribution, which has the same mathematical form as a Lorentzian profile. The case $\xi = 0$, called the Gumbel distribution, separates these two types of distinct behaviour. It is defined as the limit function when $\xi \rightarrow 0$, and has the form
\[
G(z) = \exp\left\{ - \exp \left( - \frac{z-\mu}{\sigma} \right) \right\}
\]
with $zÊ\in \Real$. Its tail decreases exponentially, giving much lower probabilities to observe very high maxima than the case $\xi>0$, but these probabilities are nowhere 0, differently from the case of a negative shape parameter. This is the limit distribution of maxima of variables from distributions like the normal (Gaussian), lognormal, gamma, exponential or the chi-squared.

\subsection{Inference and diagnostics for extremes}\label{subsec_evddiag}

The estimation of a GEV model for a time series of length $n$ starts usually with the division of the series into blocks of $k$ observations (for example, a block can be a year for a sequence of daily temperature measurements spanning several decades). From each block, we select the maximal value (in the example, the maximum temperature of each year, say $ z_1, \ldots, z_m$), and we fit the GEV model to the maxima for example by maximum likelihood (see e.g. \citealt{colesbook}) or probability-weighted moment method \citep{hoskingetal85}. The  log-likelihood, which is  derived from the cumulative distribution function  \eqref{eq:gev} by first differentiating it with respect to $z$, then computing the likelihood of the set of maxima $ z_1, \ldots, z_m$, finally taking its logarithm, takes the following form:
\begin{eqnarray} \label{eq:loglik}
\ell(\xi,\sigma,\mu) &=&  - m \log \sigma - \left(1+\frac{1}{\xi}\right) \sum_{i=1}^m \log\left(1 + \xi \frac{z_i-\mu}{\sigma}\right)  \nonumber  \\
& & -  \sum_{i=1}^m \log\left(1 + \xi \frac{z_i-\mu}{\sigma}\right)^{-1/\xi},
\end{eqnarray}
under the constraint that $1+\xi(z_i-\mu)/\sigma > 0$ for $i = 1, \ldots, m$. This function is maximized with respect to its arguments $\xi, \sigma$ and $\mu$ to obtain the best-fit estimates for the parameters. The estimates are asymptotically normal if $\xi > -0.5$ \citep{smith85}, so variance-covariance matrix and standard errors can be straightforwardly derived for the estimates. We first compute the negated second derivative matrix $-\frac{\partial^2 \ell}{\partial \theta_i \partial \theta_j}$ of the likelihood function, where the parameter vector $\theta$ denotes $\theta = (\xi, \sigma, \mu)$. We evaluate this matrix at the estimated parameter values; this is called the observed information matrix. Then the variance-covariance matrix of the estimates can be obtained by inverting the observed information matrix; the diagonal elements give the variance of the corresponding parameter estimate, off-diagonal elements provide the covariance between two estimated parameters.

The estimated GEV distribution can then be used to obtain return levels, probabilities of very high levels or distributions of maxima of longer periods. The return level $\zeta_p$ associated with the return period $1/p$ is the level that is expected to be exceeded once in every $1/p$ blocks of the same length $k$ (continuing the previous example, the $\zeta_{1/20}$ return level would be the temperature which is exceeded only once in every 20 years). This is simply the $1-p$ quantile of the GEV distribution:
\begin{equation}\label{eq:rl}
\zeta_{p} = G^{-1}(1-p) = \mu - \frac{\sigma}{\xi}\left[ 1- \left\{ - \log(1-p)\right\}^{-\xi} \right],
\end{equation}
where $G^{-1}$ denotes the inverse function of $G$. Confidence intervals for $\zeta_{\alpha}$ can be obtained by several methods: by parametric or nonparametric bootstrap, by the delta method using the variance-covariance matrix of the GEV parameters, or by profile likelihood. About the details of all these procedures, and in general about GEV modelling and  inference for extremes, \citet{colesbook} gives an excellent practical summary in its Chapter 2 and Chapter 3.
 
The selection of $k$, the number of observations in a block raises a question of bias-variance trade-off. Since the GEV is a limit distribution when $k \rightarrow \infty$, $k$ must be large enough to provide a good extreme-value approximation, otherwise the model will be poor, and yields biased parameter estimation and extrapolation. But if $k$ is too large, we can obtain only a few blocks in the series and therefore too few maxima. This implies a large variance of the parameter estimates. The choice of the block size is governed by pragmatic considerations, as in the case of annual maxima for climatic time series. In cases where the choice is not straightforward, several block size may be used to perform the analysis. Then the quality of each model is checked by diagnostic plots, and the results using the smallest block size that gives acceptable model diagnostics can be accepted as final results. 

There are two important and easy-to-use types of diagnostic plots. The first is the quantile-quantile plot, generally used in statistics to check adequacy of a fitted model instead of histograms, since histograms are sensitive to bin size choice and not adapted to detect discrepancies in the tails of the fitted distributions, exactly where FAP estimates are expected to be precise. Let $z_1, \ldots, z_m$ be a collection of block maxima,  $z_{(1)}, \ldots, z_{(m)}$ the ordered sample in increasing order, and $\hat G(z)$ the estimated GEV distribution. The quantile-quantile plot consists of the points 
\[
\left\{ \hat G^{-1} \left(\frac{i}{m+1}\right), z_{(i)} \right\}, \quad \quad \quad i = 1, \ldots, m.
\] 
If the model is good, the points should closely follow a straight line with intersect 0 and slope 1 without strong systematic deviations. This plot is a direct visual comparison of the empirical distribution function (EDF) of the data and the fitted model. The EDF is defined such that it takes the value  $\frac{i}{m}$ at the points $\{  z_{(i)} \}$; in order to avoid to have exactly 1 outside the data range, which could cause problems, we apply a small correction by using $m+1$ in the denominator. The same probability levels $\frac{i}{m+1}$ are taken by  the fitted model at $\hat G^{-1} \left(\frac{i}{m+1}\right)$. If the model is good, the EDF and $\hat G(z)$ should be close to each other. Thus, the values $\hat G^{-1} \left(\frac{i}{m+1}\right)$ and $ z_{(i)} $ should also be nearly equal, because they are the quantiles belonging to the same probability levels $\frac{i}{m+1}$ in the two distributions. 

The top row of Figure \ref{fig:gevQQ} shows two examples of quantile-quantile plots resulting from GEV fits. Both plots show acceptable fits: the grey dots representing the observed values versus the model-predicted quantiles form approximately a straight line, and no strong systematic deviations from the fit can be seen, though there is some scatter at the right end of the distribution. Another example of a general quantile-quantile plot is the right panel of Figure \ref{fig:nonnormality}, using standard Gaussian quantiles instead of the GEV, corresponding to the supposed distribution of the sample. On that plot,  non-Gaussianity can be detected as systematic deviation from the straight line. 

Whether the scatter of the right end of the distribution in Figure \ref{fig:gevQQ} is just due to natural fluctuations or to an invalid model can be better judged by the return level plot. It shows the return level function $\zeta_p$  (the inverse of the fitted distribution) against transformed probability values $\log[ -\log (1-p)]$, that is, the points
\[
\left\{ \log[ -\log (1-p)],  \zeta_p \right\},
\]
where the return level $\zeta_p$ is given by \eqref{eq:rl} for any $p$, using the fitted GEV parameters. The transformation of the probabilities is such that the Gumbel distribution would become a straight line, a heavy-tailed GEV would curve upwards above it, and a finite-tailed one would remain below it, monotonically increasing, but eventually tending to a horizontal line. 

The return level plots are complemented with confidence bands, which are indispensable in judging the quality of the fitted model. Adding the observed points 
\[
\left\{\log \left[-\log \left( 1 - \frac{i}{m+1}\right)\right],   z_{(i)}\right\}, \quad \quad \quad i = 1, \ldots, m,
\]
they usually show some scatter around the line, especially the highest ones. The confidence bands help to see how far from the fitted line they are. Despite an eventual scatter, a model is still acceptable if the points remain within the confidence bands. 

As an illustration, the return level plots corresponding to the quantile-quantile plots of Figure \ref{fig:gevQQ} are given in the bottom row. They are constructed using the same collection of maxima and the same model fit. The solid line on both graphs represents the fit, namely the return levels  $\zeta_p$ given by equation \eqref{eq:rl} against their transformed probabilities. The distribution plotted on the left-hand side is finite-tailed, the one on the right is very close to a Gumbel distribution. The confidence bands give an immediate visual impression about the uncertainty of the estimated return levels. In both cases, the fitted model is good, since all the points remain well within the confidence bands.

\subsection{Application of extreme-value theory to periodogram peaks}\label{subsec_evdper}

The classical periodogram values of an independent, identically distributed (iid) Gaussian noise evenly sampled at times $\delta t, 2\delta t, \ldots, T\delta t$, computed at the Fourier frequencies $f_j = j/(T\delta t), \; j = 1, \ldots, T/2$, are distributed independently as standard exponential variables (or equivalently, apart from a normalizing factor, as $\chi^2_2$ variables). Consequently, the extreme-value limit for its maximum is the Gumbel distribution. When the iid sequence is non-Gaussian, the classical periodogram at the Fourier frequencies is neither independent nor uncorrelated, though asymptotically, the values at any set of frequencies form an  iid standard exponential vector when $T\rightarrow\infty$ \citep[Proposition 10.3.2]{brockwelldavis06}. \citet{davismikosch99} have proven the Gumbel limit also for the maximum at the Fourier frequencies for the classical periodogram of non-Gaussian iid sequences, when the mean of the sequence is zero and the variance is finite. \citet{linliu09} generalizes this further to continuous transformations of linear processes, which satisfy several conditions on the tail decay and the moments of the underlying noise process, in addition to restrictions on the dependence structure and summability conditions on the linear process. These works allow to test the hypothesis of periodicity against more complex dependent non-Gaussian processes, if the observed sequence is regularly sampled, and the (classical) periodogram is computed at the Fourier frequencies.

In the evenly sampled case summarized above, increasing the number $T$ of observations leaving $\delta t$ fixed leads directly to the increase of the number of Fourier frequencies in a fixed frequency interval, and an asymptotic limit, where extreme-value theory works, follows naturally. When turning to oversampled periodograms of irregularly observed time series, the situation becomes more complex. We observe $N \ll T$ data points, so we have a majority of missing values in the time domain, and we compute the periodogram on a frequency grid that is oversampled by a factor $K$ with respect to the Fourier grid. The ensuing degeneracy (see Section \ref{subsec_fou}) in the joint distribution of the periodogram makes the extreme-value limits non-trivial. As the existence of the limiting GEV distribution is linked to only an asymptotically weakening dependence at extreme levels, a strong dependence in a finite case does not exclude the existence of such a limit. It can be assured  if we find particular specifications for the relative rates of $T \rightarrow\infty$, $N \rightarrow\infty$ and $n \rightarrow\infty$, so that dependency and degeneracy in the spectrum decreases. Such a specification is given by $K \rightarrow 1$, $T \rightarrow \infty, \, N \rightarrow \infty$ such that $N/T \rightarrow 1$. In this case, the proportion of the missing values in time-domain decreases towards 0, and the refined frequency grid, as determined by the oversampling factor $K$ decreasing to 1, tends to the Fourier grid. The limit is then the evenly sampled case with the Gumbel distribution.  

Hence, several facts motivate testing simple extreme-value methods for the assessment of significance of periodogram peaks: (1) the existence of a Gumbel limit for the classical periodogram maxima at the Fourier frequencies for a large class of time series under even sampling; (2) the possibility to define time sampling and missing value patterns such that in a limit $N \rightarrow \infty$ we approach an evenly sampled case; (3) the general nature of the extreme-value family as the limiting distribution of almost all continuous distributions under broad dependence conditions; and (4) this limit, if exists, can be estimated in a straightforward way for the null hypothesis $H_0$. Estimation or knowledge of the number of independent frequencies and stringent assumptions on the marginal distribution of the periodogram are not necessary anymore. 

For our procedure, we assume therefore the existence of a GEV limit for the maxima of periodograms. Moreover, we assume that in the finite cases we are dealing with, the GEV provides a sufficiently good approximation to the true distribution of the maximum. The multitude of methods used in astronomy, each of which has its particular distribution of periodogram, and the degeneracy in the probability distribution of the periodogram suggests the use of the general GEV family with $\xi \in \Real$, instead of the Gumbel subfamily with $\xi = 0$ found for the classical periodograms. A further motivation is that any asymptotic theoretical distribution of the periodograms can be wrong in the high quantiles, and strongly deviate from the exponential tail. The generalization from $\xi = 0$ to $\xi \in \Real$ admits all continuous distributions, not only the exponentially-tailed ones.  However, as for all tail estimation, the estimates are very sensitive to these underlying assumptions. Thus, the quality of the GEV approximation must always be checked with the diagnostic plots presented in Section \ref{subsec_evddiag}.

\section{Estimation of the FAP}\label{sec_proc}

\subsection{Procedure}\label{subsec_proc}

\begin{description}

\item[\textbf{1. Bootstrap of the original time series.$\quad\quad\quad\quad\quad\quad$}]
In order to generate noise sequences under $H_0$, we resample the original observations $X_1, \ldots, X_N$ with replacement and with equal probabilities (nonparametric bootstrap) $R$ times, using the same observational epochs. Thus, we create $R$ repetitions $X_{j_1}^{(r)}, \ldots, X_{j_N}^{(r)}$ of a white noise series with approximately the same marginal distribution. 
\item[\textbf{2. Maxima of partial periodograms.$\quad\quad\quad\quad\quad\quad$}]
From the frequency grid  $\mathcal{F}$, we randomly select $L$ frequency intervals of $K$ consecutive frequencies, where $K$ is the oversampling factor, and $L$ is chosen large enough to provide a good extreme-value approximation. This is equivalent to a random draw of $L$ central frequencies $f_{j_1}^{(r)}, \ldots, f_{j_L}^{(r)}$ with equal probabilities,  and, around each, taking a frequency interval containing $K$ consecutive grid frequencies ($K$ is the oversampling factor with respect to the Fourier grid). The periodogram must be calculated only at these frequencies, and only the maximum of each partial periodogram is needed. The output of this step is thus a sample of $R$ maxima of partial periodograms of white noise sequences, which have distributions similar to the original observations. 
\item[\textbf{3. GEV modelling of the partial maxima.$\quad\quad\quad\quad$}] 
Fit an extreme-value model $G(z; \xi, \sigma, \mu)$ to the $R$ maxima, maximizing the log-likelihood \eqref{eq:loglik}, to obtain estimates $\hat \xi, \hat \sigma$ and $\hat \mu$ for its parameters, and compute the inverse of the observed information matrix for their uncertainty estimates. Use diagnostic plots to check the quality of the fit.
\item[\textbf{4. Extrapolation for the complete periodogram.}] 
If the quality of the fit is sufficiently good, use the estimated $\hat G(z; \hat \xi, \hat \sigma, \hat \mu)$ to find levels corresponding to the desired FAP values. Give confidence intervals of these levels.

\end{description}

\subsection{Reasoning behind the steps}\label{subsec_arg}

\begin{description}

\item[\textbf{1. Bootstrap.}] We propose here nonparametric bootstrap as an alternative to assume a specific parametric form for the noise. Fixing a parametric distribution for it is equivalent to fixing a specific subfamily of the GEV distribution, corresponding to either $\xi < 0$ (finite tail), $\xi = 0$ (exponential tail) or $\xi >0$ (heavy tail). For example, assuming Gaussianity of the observations or a $\chi^2_2$ distribution for the periodogram implies at once an exponential tail with $\xi = 0$. This is a very important restriction, with crucial impact on the very high quantiles necessary for FAP. Also, not only the estimated parameters (and quantile levels) are biased in such a case, but the variance of the parameter estimates is too small. It does not reflect our (real) uncertainty about the tail of the observations or the periodogram, and  gives a false measure of how sure we can be about an eventual detection of periodicity.

In addition, to check the null hypothesis of the observed sequence being a white noise, we should admit the possibility of deviating from any presumed distribution. In the situation of Figure \ref{fig:nonnormality}, it is clear that if this sequence is indeed noise, then it cannot come from a Gaussian distribution. Generating white noise from a Gaussian distribution would produce on average lower periodogram maxima than another, non-Gaussian distribution that is closer to the observed one, since the observed upper tail seems to be heavier than a Gaussian distribution. Thus, Gaussian simulations would overestimate the significance of the maximum of the periodogram of the observations. 
  
The nonparametric bootstrap is a flexible alternative to a parametric bootstrap, in order to avoid such restrictions on the tails. It corresponds to using the empirical distribution function $\tilde F(x) = \frac{1}{N} \sum_{i = 1}^N I (x > X_i)$ instead of the true unknown distribution $F(x)$ to generate a noise sequence, i.e., an independent, identically distributed random sample of the distribution of the data. The parametric and the nonparametric bootstrap were compared in simulations;  there was no perceptible difference in the estimated GEV parameters for the two different types of simulations. 
\item[\textbf{2. Maxima of partial periodograms.}] \hfill
{Two different} \\ aims motivate the particular way of the procedure to select "blocks", subsets from periodograms of which we take the maxima. The principal goal is to decrease the computational loads due to a bootstrap. At the same time, the reduced frequency set must reflect the fundamental characteristics of a full periodogram: the long-range dependence manifested in the non-vanishing correlations between sinusoids of two distant frequencies, and the short-range dependencies due to the spectral leakage. The random selection of central frequencies throughout the whole range $(0, f_{\max}]$ provides a sample which is representative of the long-range dependence, whereas taking intervals of width equal to the oversampling factor $K$ accounts for the effects of spectral leakage.  Maxima of such partial periodograms carry information on both kinds of dependency, and the GEV fit of step 3. will reflect these, but a careful assessment of model quality is required to enable extrapolation. 

\item[\textbf{3. GEV modelling.}] The fitting can be done in several ways, for example by the method of maximum likelihood \citet{smith85,colesbook}. The parameter estimates, if the true $\xi>-0.5$, follow an asymptotic multivariate normal distribution (similar to the usual asymptotics of regular likelihood estimates), so inference is straightforward for the estimated model. Due to the strong dependence and the degeneracy present in the periodogram, and to check whether the number of bootstrap repetitions $R$ and the size of the partial periodograms $KL$ provide a sufficiently good extreme-value approximation, it is necessary to assess model quality by the diagnostic plots described in Section \ref{subsec_evddiag}.
\item[\textbf{4. Extrapolation.}] The model can be used for extrapolation only if the diagnostics have proven the model acceptable. This consists of finding quantile levels of the full periodogram, based on the extreme-value modelling of the partial periodogram. A level $\zeta_{\alpha}$ corresponding to FAP $= \alpha$ in the full periodogram is exceeded only once in $1/\alpha$ complete periodograms (in numbers, FAP = 0.01 means that out of a hundred periodograms, we can expect on average only one where the maximum exceeds $\zeta_{\alpha}$). As the complete periodogram contains $n/(KL)$ times more test frequencies than the subsets used for estimation, this corresponds to one exceedance in  $n/(\alpha KL)$ partial periodograms. Using Eq. \eqref{eq:rl}, we can then compute the desired level as $\zeta_{\alpha} = \hat G^{-1}(1-[\alpha KL]/n)$, and can add also confidence intervals.

Though using only the maxima of a reduced frequency set for the estimation of the GEV parameters alleviates the problems due to the time-requirements of the bootstrap, the reduction implies that we need to extrapolate in order to give return levels of maxima of the complete periodogram. If the used frequency set is by far smaller than the complete periodogram, the extrapolation must reach to levels far beyond those used in the fit, and the estimated return levels will have a large uncertainty. A frequency set size closer to the complete periodogram provides more reliable extrapolation. Thus, there is a trade-off between the computational load and the necessary range of extrapolation. We used simulations with various deterministic patterns, error distributions and signal-to-noise ratios to show that the estimates are remarkably stable down to frequency set size $L =100$ and repetition number $R$ around 200-400.  The choice of $R$ and $L$ can also be checked by the above mentioned diagnostic plots: for bad model fits, increasing $L$ is a possible remedy, that is, using larger partial periodograms, and if the observed points scatter too much around the fitted line, this may be improved by increasing R, that is, using a larger number of bootstrapped noise sequences. 

\end{description}

\section{Results}\label{sec_res}

\begin{figure}
\begin{center}
\includegraphics[scale=.6]{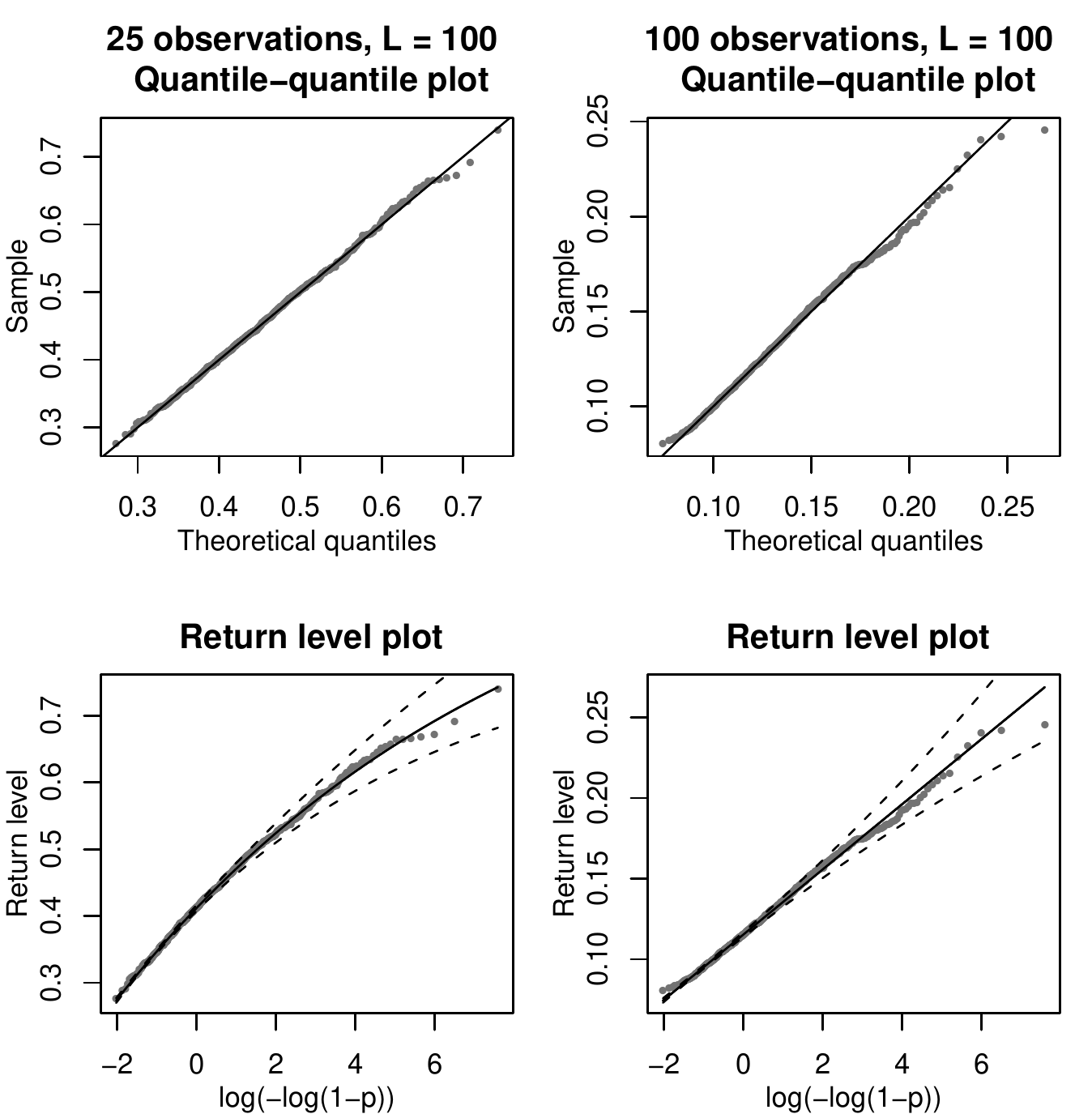}
\caption{Upper panels: quantile-quantile plots for two GEV fits for the maxima of 1000 noise sequences generated by bootstrap, for 25 and for 100 night-time observations (left and right panel, respectively).  In both cases, the bootstrapped original time series was the sinusoidal simulation with SNR = 1, and $L =100$ random frequency intervals were used. Bottom row: the return level plots for the same data as on the panels above. Dots correspond to the observed values against the transformed empirical probabilities $\log \left[-\log \left( 1 - \frac{i}{m+1}\right)\right]$, solid lines to the fitted model, and dashed lines to a 95\% confidence interval for the fitted return level curve based on asymptotic normality of the maximum likelihood estimator.}
\label{fig:gevQQ}       
\end{center}
\end{figure}

\subsection{Simulations}\label{subsec_sim}

The performance of the procedure was assessed using two light curve patterns, a sine-wave of the form $g(t) = A_{\sin,i} \sin(f_{\sin} t) $  and a broken-line model for detached eclipsing binaries. For the sinusoid, the light curve parameters were $f_{\sin} = 3.379865 \; d^{-1}$ and three different amplitudes $A_{\sin,1} = 0.15, A_{\sin,2} = 0.05$ and $A_{\sin,3} = 0.025$ mag. For the eclipsing binary, $f_{\rm ecl} = 0.4243146 \; d^{-1}$ was used, with the depth of the primary minimum equal to $A_{{\rm ecl}, i} = 2, 1.33, 1, 0.67, 0.33$ and $0.167$ mag, and the ratio between the depth of the secondary and the primary minima fixed to 0.375.

To both variability patterns, we added random noise generated from Gaussian distributions with time-varying variance, yielding the model $Y_i = g(t_i) + \epsilon_i$ with $\epsilon_i \sim {\mathcal N}(0, \sigma_i^2).$
The time-varying standard deviations $\sigma_i$ were themselves random, based on a random gamma-distributed variable at each point, yielding an average $\sigma_i$ of 0.05 mag (the approximate signal-to-noise ratios of the three sine-wave simulations were thus SNR$_{\sin,i} =A_{\sin,i} / \bar \sigma = 3,1$ and $0.5$). The deviations added to the sine-wave and the eclipsing binary simulations at point $i$ were then generated from the Gaussian $ {\mathcal N}(0, \sigma_i^2).$ 

Epochs of observations were chosen on a time grid of 0.005 day (about 10 minutes) with a total span of 25 days. Imitating random nighttime observations during 4 hours each night, two different sequences of epochs were randomly uniformly selected from 25 nightly 4-hour periods. One consisted of $N=100$ observations, the other, of $N=25$.  The input data for the period search were the two sequences of epochs, the nine noisy light curves each sampled at both cadences as observed values and the random gamma variables as error bars, yielding in total 18 different time series. 

The time grid parameters led to an upper frequency detection limit $f_{\max} = 100 d^{-1}$, a Fourier frequency set of ${\mathcal F}_F = \{0.04, 0.08, \ldots, 100 \} d^{-1}$ and a corresponding peak width of 0.04 $d^{-1}$ due to leakage. An oversampling factor $K=16$ was used, providing a test frequency grid ${\mathcal F} = \{0.0025, 0.005, \ldots, 100 \}\; d^{-1}$ with $n=40000$ test frequencies. The generalized Lomb-Scargle method was performed on all 18 time series using these test frequencies, once without weighting, and once with weights defined by the normalized inverse squared error bars. The grey spikes in Figure \ref{fig:sinPGQ} show the resulting periodograms from the non-weighted version for the six sinusoidal light curves with Gaussian errors. 

The procedure presented in Section \ref{sec_proc} was performed on the 18 light curves. In order to check the stability and the variance of the estimates as a function of the number of bootstrap repetitions $R$ and the number $L$ of the test frequency intervals, we applied all pairwise combinations of $L =  50, 100, 200, 300,400,500$ and $R = 200,400,600,800,1000$ for each of the 18 simulated light curves. We calculated also the complete periodogram for 2000 independent bootstrap noise sequences for each of the simulated light curves, in order to compare the model-based and the empirical high quantiles, corresponding to selected FAP levels.

\subsection{Model fits and diagnostics}\label{subsec_diag}

Each combination of number of repetitions and number of random frequency intervals $(R, L)$  on each simulation yielded a sample of $R$ maxima of a partial periodogram of size $KL$, which were then fitted with the GEV model. The quality of all models was checked by the return level and quantile-quantile plots described in Section \ref{subsec_evd}. In general, the models are acceptable and can be used for extrapolation, though some small deviations could be found in many plots, as shown in Figure \ref{fig:gevQQ}. The model quality on average is somewhat worse for the shorter time series with $N=25$ than for time series with $N=100$, but no other systematic difference can be seen with respect to $R$ or $L$. The high end of the quantile-quantile plots often deviates slightly downward, implying that the observed periodogram maxima are stochastically smaller than the model estimates. The same effect can be remarked also on the return level plots: the theoretical curve is above the points corresponding to the observations. This leads to a conservative error in the significance assessment: the model-based estimated quantiles will be a little higher, and an observed peak will be found somewhat less significant.  The converse error is very rare, which implies a lower risk of false periodicity identifications.

\subsection{Plausibility of quantile estimates and stability}\label{subsec_bias}

\begin{figure*}
\begin{center}
\includegraphics[scale=.76]{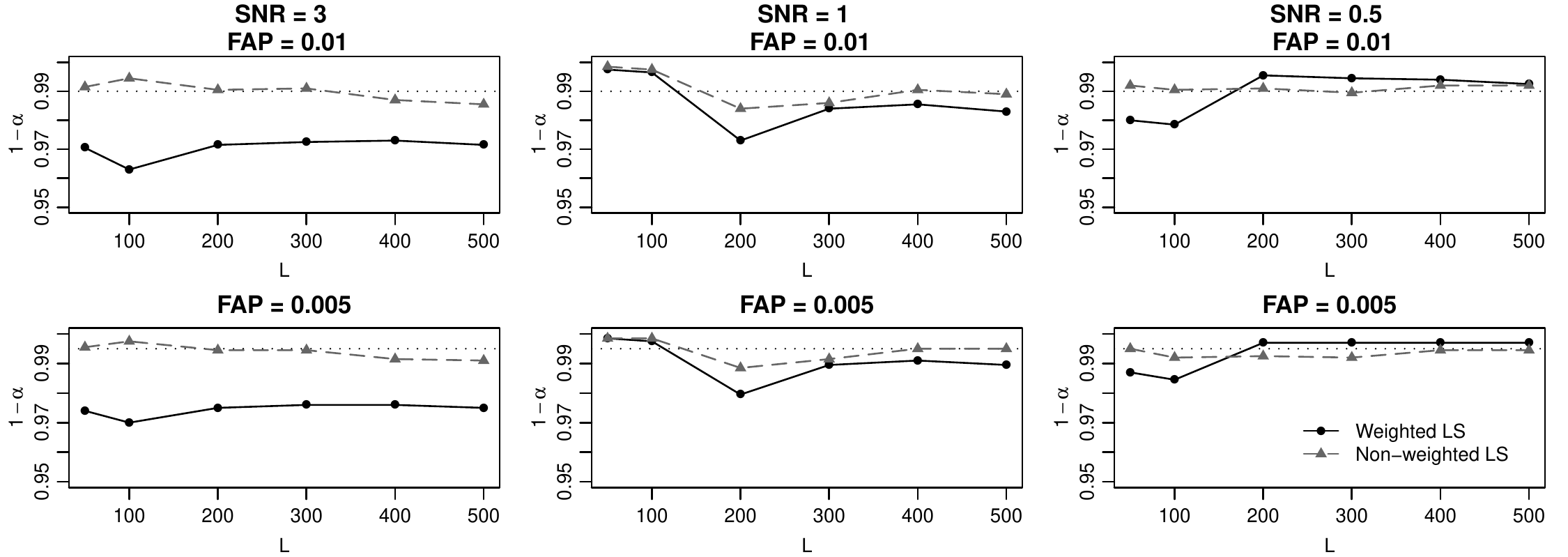}
\caption{The proportion of full periodogram maxima in 2000 that exceed the estimated high quantiles for FAP $= 0.01$ (top row) and  FAP $= 0.005$ (bottom row), as a function of $L$, for time series length $N=25$, with $R = 1000$. Dashed grey line with triangles show the results using non-weighted generalized Lomb-Scargle, solid black lines with dots is using the weighted version. The dotted lines denote the FAP levels. The left panels refer to SNR$=3$, the middle panels, to SNR$=1$, and the right panels, to SNR$=0.5$. }
\label{fig:stabL25}       
\end{center}
\end{figure*}

\begin{figure*}
\begin{center}
\includegraphics[scale=.76]{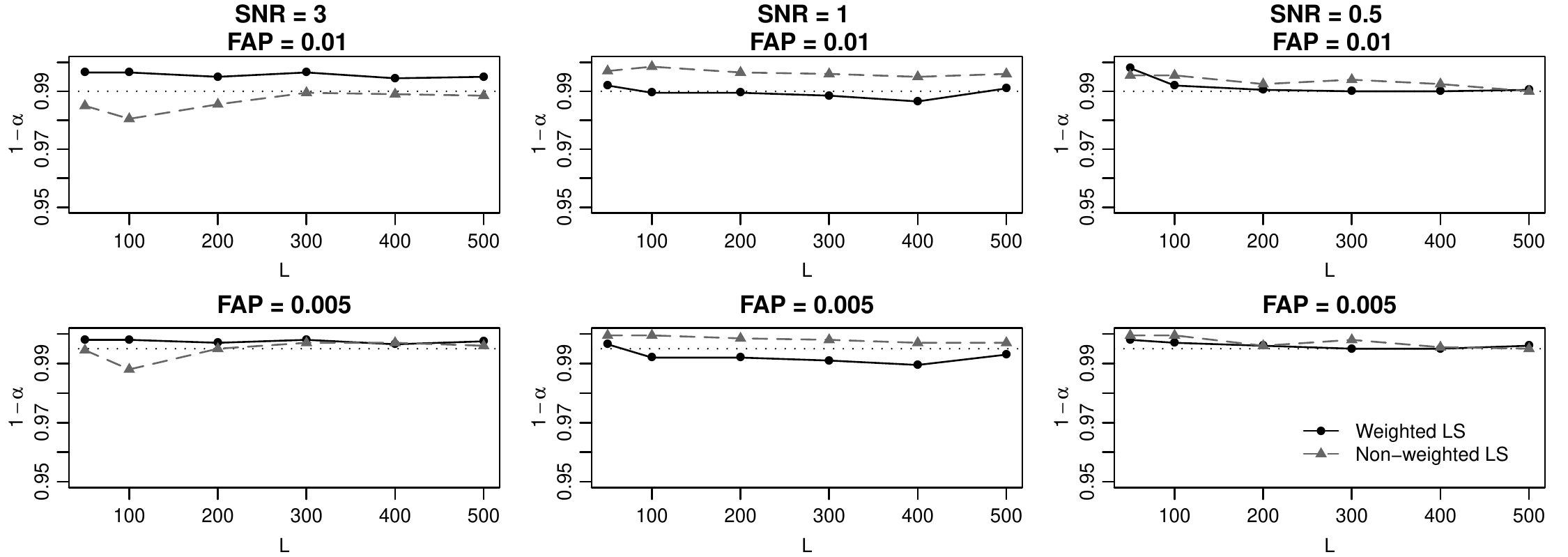}
\caption{The proportion of full periodogram maxima in 2000 that exceed the estimated high quantiles as a function of $L$, for time series length $N=100$. The symbols and the SNR-FAP combinations are the same as in Figure \ref{fig:stabL25}. }
\label{fig:stabL100}       
\end{center}
\end{figure*}

After the inspection of the diagnostic plots, all GEV models obtained for the 18 simulations with all $(R,L)$ combinations were used to estimate quantile levels $\hat \zeta_{\alpha}$ belonging to fixed FAP levels $\alpha=0.01$ and $0.005$. In order to check their plausibility, we created 2000 bootstrapped white noise sequences from each of the 18 simulated time series. The complete periodograms of every repetitions were computed, and the maxima of these selected. For a FAP equal to $\alpha$, the proportion $\hat\alpha$ of the peaks exceeding $\hat \zeta_{\alpha}$ was calculated, and compared to $\alpha$. 

Figures \ref{fig:stabL25} and \ref{fig:stabL100} show the results for the six sine-wave simulations with Gaussian noise, $1 - \hat\alpha$ as a function of $L$ for the short and the long time series, respectively. The agreement between the theoretical $1-\alpha$ and the empirical $1-\hat\alpha$ is visibly much better for the non-weighted method than for the weighted for the short time series. A possible explanation is that though weighting for least squares regression improves on the estimates only when the true model is fitted, namely, a sine-wave signal plus Gaussian noise with known variances, under $H_0$ we fit only a constant plus , and weighting with the inverse variances loses its optimality property. An improvement with decreasing signal-to-noise ratio, when the deviations from normality become smaller, is visible in Figure \ref{fig:stabL25}, which supports this explanation.

Small test frequency set sizes ($L=50$ or 100) cause instability of the empirical exceedance proportions $1-\hat \alpha$, especially for the short time series. With $N=25$ and for the weighted period search version, $L>200$ seems necessary. Above this,  $\hat \alpha$ is stable, and approximates well the FAP level
  $ \alpha$ even in the case of short time series. For the longer time series with $N = 100$, the stability is remarkable, and the approximation is in general good.
   
\begin{figure*}
\begin{center}
\includegraphics[scale=.28]{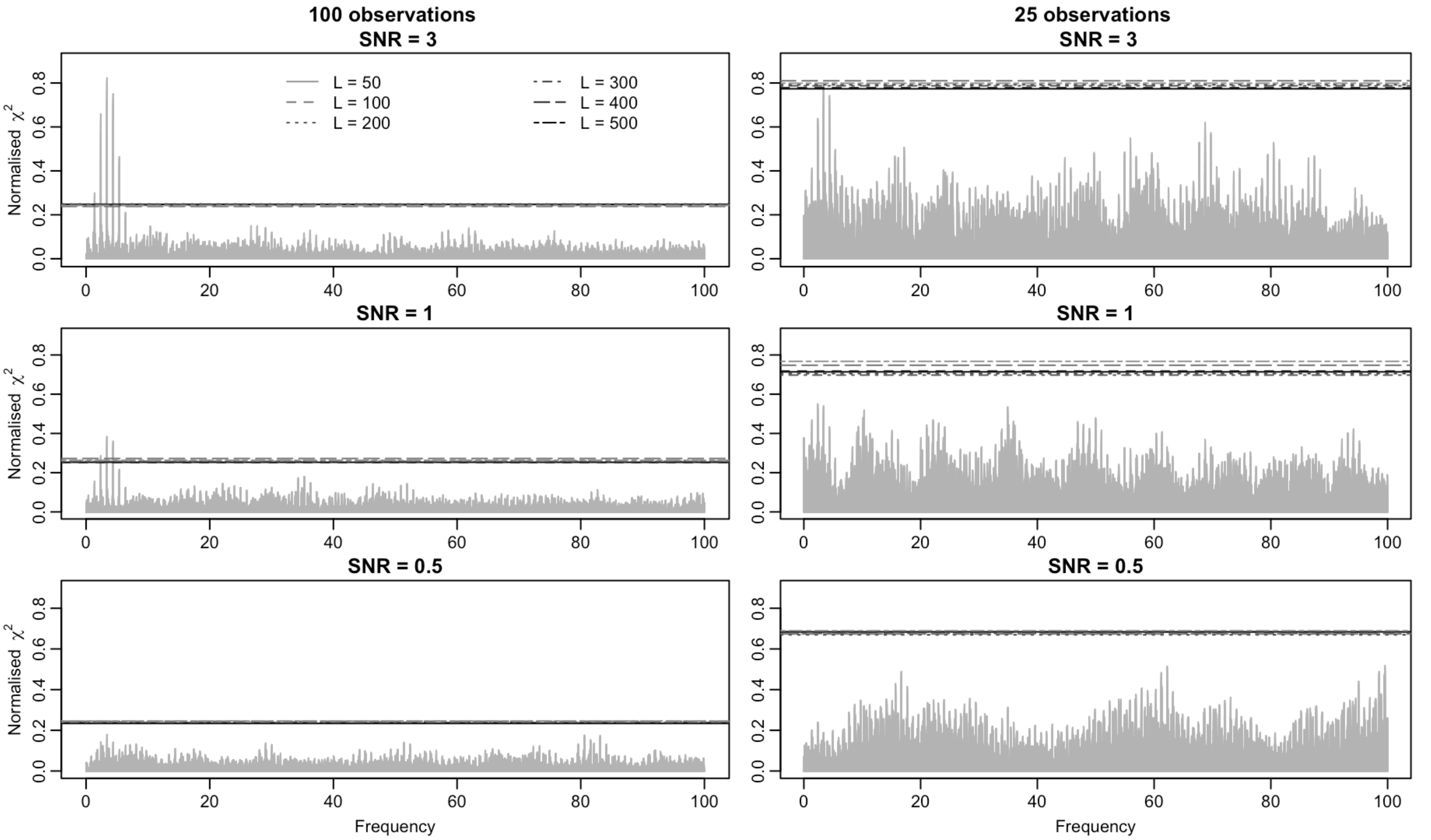}
\caption{Stability of the estimated 0.99 quantile as a function of the number of test frequencies for the sinusoidal simulations with Gaussian errors, $N = 100$ (left column) and $N=25$ (right column). The grey spikes show the periodograms of the simulated noisy sinewave signals, the horizontal lines are the quantile levels estimated from fitted GEV models. The different types of the lines correspond to different numbers of test frequency intervals $L$, and are the same for all panels, shown in the legend in the top left panel. For all plots, $R = 1000$ and the period search method is the non-weighted generalized Lomb-Scargle.}
\label{fig:sinPGQ}       
\end{center}
\end{figure*}

\begin{figure*}
\begin{center}
\includegraphics[scale=.7]{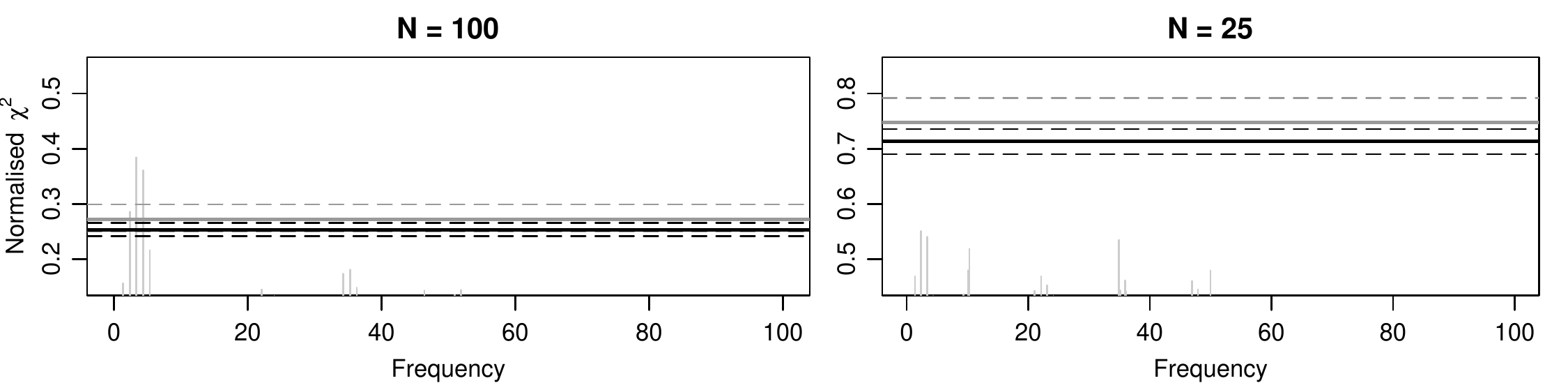}
\caption{95\% confidence intervals of the estimated 0.99 quantile for the SNR = 1 case of the sinusoidal simulations with Gaussian errors. The plots are enlarged versions of the two panels of the middle row of Figure \ref{fig:sinPGQ}. For visibility, quantiles for only two values of $L$ are plotted ($L = 100$ with grey, $L = 500$ with black); the thick solid lines correspond to the quantile estimates, the dashed ones to 95\% confidence intervals from bootstrap.
}
\label{fig:enlargedSinPGQ}       
\end{center}
\end{figure*}

The plausibility of the estimates can be easily seen if they are plotted against the periodograms of the time series. In Figure \ref{fig:sinPGQ}, the decrease of the signal below detection level can be clearly observed, as the SNR decreases. The left panels show the time series of the sinusoidal light curve with Gaussian noise with $N=100$ observations, the right panels the same light curve--noise combinations with $N=25$. The levels predicted by the procedure are presented as horizontal lines with different line types for different $L$ values. These levels reflect well the judgment based on the aspect of the whole periodogram. In the cases of SNR $=$ 3 and 1 with $N=100$ observations, the presence of a signal is obvious because of the absence of other comparable peaks, and accordingly, the plotted FAP $=0.01$ levels pass well below the peaks. In the case of the weakest signal with $N=100$, the periodogram exhibits another group of peaks of comparable height beside the correct one. The signal is undetectable based on the noisy data. The quantiles, in agreement with the impression of non-significance, appear here well above the peaks, and indicates that neither of the peaks is significant. 

\begin{figure*}
\begin{center}
\includegraphics[scale=.77]{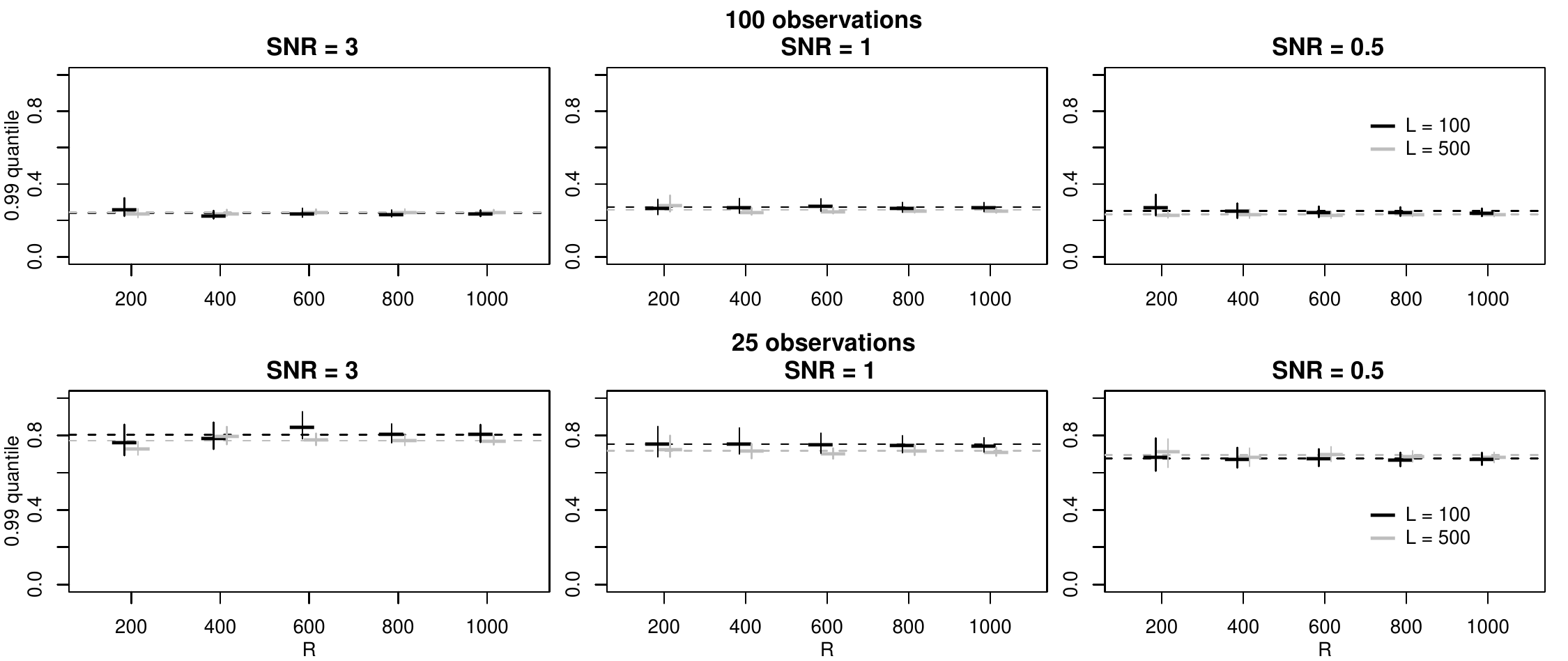}
\caption{Stability of the estimated quantile corresponding to FAP $ = 0.01$ as a function of the number of bootstrap repetitions, for number of test frequency intervals $L = 100$ (black) and 500 (grey). 95\% confidence levels based on nonparametric bootstrap are plotted as vertical bars. The plotting positions are slightly shifted horizontally only in order to avoid overlapping marks.}
\label{fig:stabrep}       
\end{center}
\end{figure*}

With less data ($N = 25$), the same plausibility can be observed. So scarce sampling makes even a signal of SNR $=3$ barely detectable. A careful judgment says that there is very likely a signal, but the maximum of the periodogram does not exceed very much the secondary peak, and there may be a weak probability that a noise sequence produced this periodogram. Accordingly, the estimated quantiles for FAP $=0.01$ hover around the highest peak: its statistical significance is around 0.01, that is, there is $\sim 1$\% probability that a noise sequence produces such a peak. For the two other cases with smaller SNR, the signal is not detectable: for SNR $=1$, there are many other peaks with comparable size, and for SNR$=0.5$, the signal is lost, and the spectrum is dominated by peaks solely due to noise.

The plausibility for the eclipsing binary-like time series is very similar, with the only difference that the period search method finds the double of the frequency, not the correct frequency. The generalized Lomb-Scargle method does so very often for eclipsing binaries, due to their two more or less equally-spaced minima in the light curve. The significance of a peak present in the periodogram is assessed in the same way and with the same plausibility of the results for the double frequency. When applied to real data, a further step of plotting the folded light curve can decide whether the analysed star is an eclipsing binary or not, and whether the correct or the double frequency was found.

The stability with respect to the number of test frequency intervals $L$ can be observed in Figure \ref{fig:sinPGQ}, too. In the left panels with $N=100$, all estimated quantile levels corresponding to different numbers $L$ of test frequency intervals are almost indistinguishably close together. Due to the scarce data, the estimated levels are more scattered in the right panels showing the $N=25$ cases, but the agreement is still quite good, and conveys reasonable judgments about the significances. The agreement between estimates with different $L$ values can be even better appreciated in Figure \ref{fig:enlargedSinPGQ}. This shows an enlarged picture of the estimated quantiles of the simulation with SNR $ = 1$ and $R = 1000$ (middle panels of Figure \ref{fig:sinPGQ}),  together with bootstrap-based confidence intervals, offering a better opportunity to judge their stability. For the sake of visibility, only $L=100$ and $L = 500$ are plotted. The overlap of confidence intervals confirms the stability of the quantile estimates in a broad range of frequency set sizes.

\subsection{Stability with respect to $R$}\label{subsec_stabR}

The weak  dependence of the estimated high quantiles on the number $R$ of bootstrap repetitions can be seen in Figure \ref{fig:stabrep}. The upper row shows the estimated 0.99-quantiles for the complete periodogram of the sinusoidal signals with Gaussian noise with 100 observations, the bottom row shows them for $N = 25$. In the former case, $R$ as low as 200 can be combined with a number $L$ of test frequency intervals as low as 100 giving stable and reliable estimates, though the confidence intervals (here, those originating from bootstrap) are larger with such small values than with $L = 500$ or $R = 1000$. For a scarcely sampled time series, the estimates are more varying according to $L$ or $R$, but over the whole range, they agree with each other within the confidence intervals. The stability of the estimates with respect to $R$ and $L$ allows a strong reduction of computational time, preserving at the same time the quality of the significance assessment.


\section{Candidate multi-mode RR Lyrae stars from SDSS}\label{sec_rr}

\begin{figure*}
\begin{center}
\includegraphics[scale=.66]{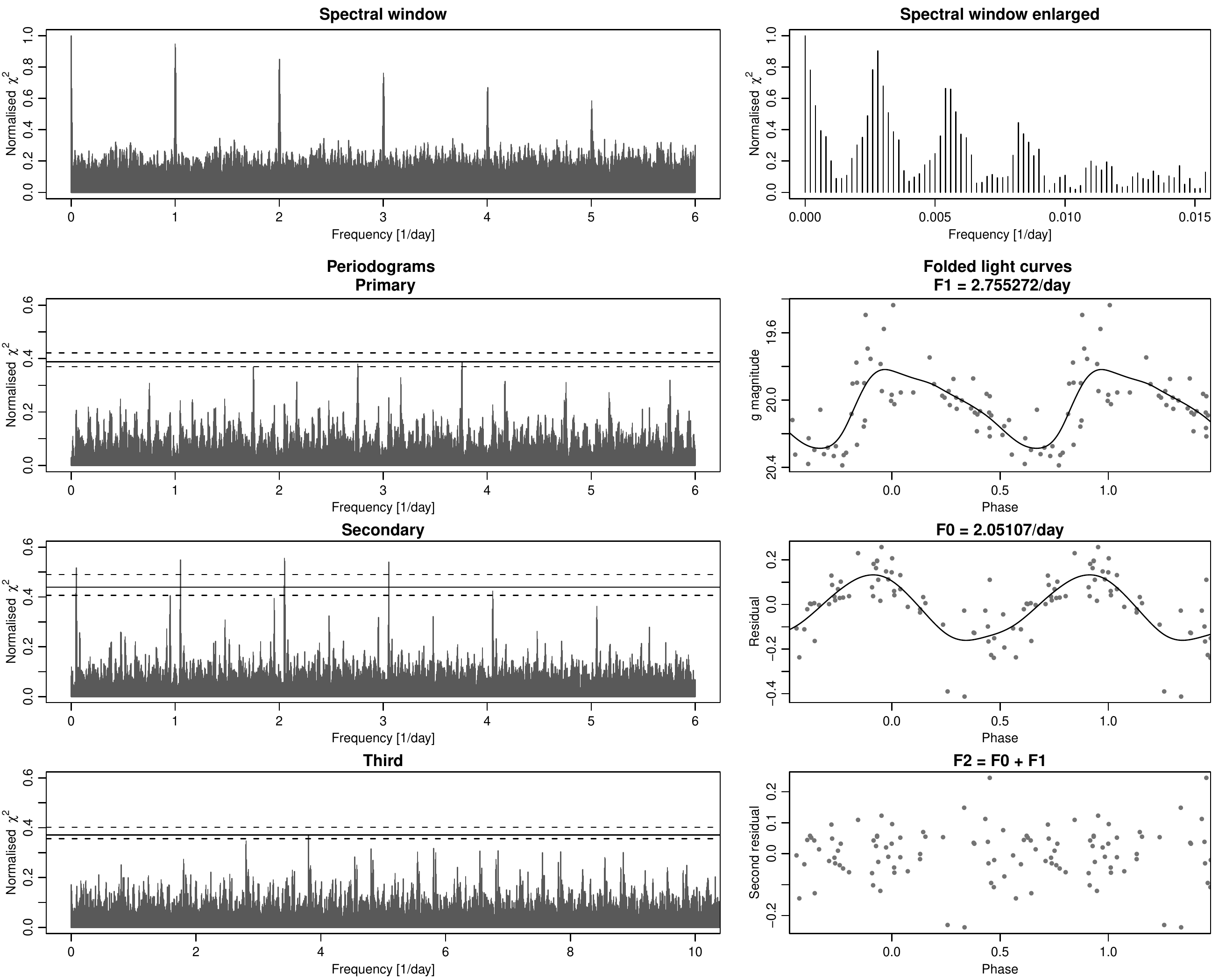}
\caption{Spectral window (top left), enlarged spectral window around 0 (top right), periodograms of $g$-band observations and of residual time series after two successive pre-whitening (left panels, second to fourth row) and the corresponding folded light curves (right panels, second to fourth row) for star  538812 (J231332.19-010746.2). On the periodograms, the FAP levels of 0.01 are plotted as solid black lines, together with their 95\% bootstrap confidence intervals.}
\label{fig:538812}       
\end{center}
\end{figure*}

\subsection{Data}\label{subsec_data}

We give two examples of the use of the proposed procedure, namely, two variable stars from the Sloan Digital Sky Survey (SDSS) Stripe 82\footnote{http://www.astro.washington.edu/users/ivezic/sdss/catalogs/ S82variables.html} \citep{ivezicetal07}. The SDSS provides five-band ($u$, $g$, $r$, $i$ and $z$) photometry of more than 11000 deg$^2$ of the sky. A region along the celestial equator, called Stripe 82, was imaged repeatedly during a 10 year long period, providing 5 simultaneous time series of up to $\sim\! \! 100$ data points per object until Data Release 7 \citep{abazajianetal09}. RR Lyrae and high-amplitude $\delta$ Scuti stars were identified in this data set by \citet{sesaretal10} and \citet{suvegesetal12b}. The classification using principal component analysis presented in the latter yielded a number of candidate RR Lyrae stars. Among these, there were many rejected because of a noisy light curve or a nonsignificant primary frequency. We reconsider two of these stars here, a double-mode candidate and one with a surprisingly high secondary frequency.

All the periodograms were computed by the non-weighted generalized Lomb-Scargle procedure between $[0, 6]  \; d^{-1}$ for the double-mode candidate or $[0, 40]  \; d^{-1}$ for the secondary periodogram for the high-frequency mode candidate, using a resolution of $0.0001 d^{-1}$. This grid choice gave for both stars an oversampling factor $K \approx 12$, which was checked by plotting the spectral window. Candidate periodic signals were removed from a time series of either observed magnitudes or residuals by fitting cyclic cubic splines to the folded light curve, because this is able to remove all harmonics of that frequency simultaneously at the expense of a milder decrease of degrees of freedom than a Fourier harmonic series. A range of smoothing parameters was tried for all light curves, among them one based on generalized cross-validation (GCV, see e.g. \citealt{ruppertwandcarrollbook}). In majority, the GCV criterion produced a reasonable smooth light curve. In the case of overfitting, the smoothing parameter was adjusted manually. The procedure proposed in Section \ref{sec_proc} was performed for all observed and pre-whitened light curves, using $R = 500$, combined with $L = 200$ or $L=800$ for periodograms with $[0, 6]$ or $[0, 40]  \; d^{-1}$, respectively. Return levels for the complete spectrum, corresponding to FAP $=0.05$ and 0.01, that is, the 0.95- and 0.99-quantiles were then calculated from a GEV model fitted to the 500 bootstrap periodogram maxima, and the decision about the significance of the periodicity was obtained by comparing the peak in the observed sequence to the estimated level. 
 
\subsection{A double-mode candidate}\label{subsec_rrd}

The spectral window and primary and residual $g$-band periodograms of the star 538812 (J231332.19-010746.2) are shown in the left panels of Figure \ref{fig:538812}. The spectral window exhibits slowly decaying daily alias peaks. Its enlarged version in the top right panel shows in addition strong yearly aliasing and an oversampling factor $K = 13$. The star has a weakly significant $(0.01 <$ FAP $< 0.05)$ primary peak at the frequency $2.755272 \; d^{-1}$ in $g$. Period search on the other bands supports the existence of a signal at this period: the periodogram maximum in $u$ falls at a yearly alias of this frequency, and there is a prominent peak here in the other three bands as well, though those are not maxima. There are apparently several other periodic signals in the other bands, suggesting a multi-periodic nature. Pre-whitening all bands with the spline smoother yields a very significant peak in the residual spectrum at $2.05107 \; d^{-1}$, providing the fundamental frequency $f_0$ of the star. The proportion between the two frequencies is 0.7444, which corresponds perfectly to the Petersen diagram of double-mode RR Lyrae stars. Performing pre-whitening with the same technique on the other bands yields similarly significant periodicities in $r$ and $i$, but no evident signal in either $u$ or $z$, probably due to the higher noise levels in these bands and to the faintness ($\sim$20 mag in $g$) of the star. The removal of the second cycle does not yield any further significant new frequency, apart from an alias of $f_0 + f_1$, and since the residual degrees of freedom decrease by a value around 10 or more with each pre-whitening, the data set size ($N = 56$) does not allow any further meaningful investigations.

The very weak significance of the primary peak is due to the strong secondary frequency, which causes large residual noise in the folded light curve, and hence a small relative decrease of $\chi^2$ in the primary periodogram. The found frequencies, the position of this star on the Petersen diagram and on the $(u-g,g-r)$ and period-amplitude diagram, together with the results of a principal component analysis similar to \citet{suvegesetal12b}, confirms its double-mode nature despite the only weakly significant frequency in the primary spectrum.

\subsection{A candidate with a high-frequency secondary mode}\label{subsec_rrm}

\begin{figure*}
\begin{center}
\includegraphics[scale=.66]{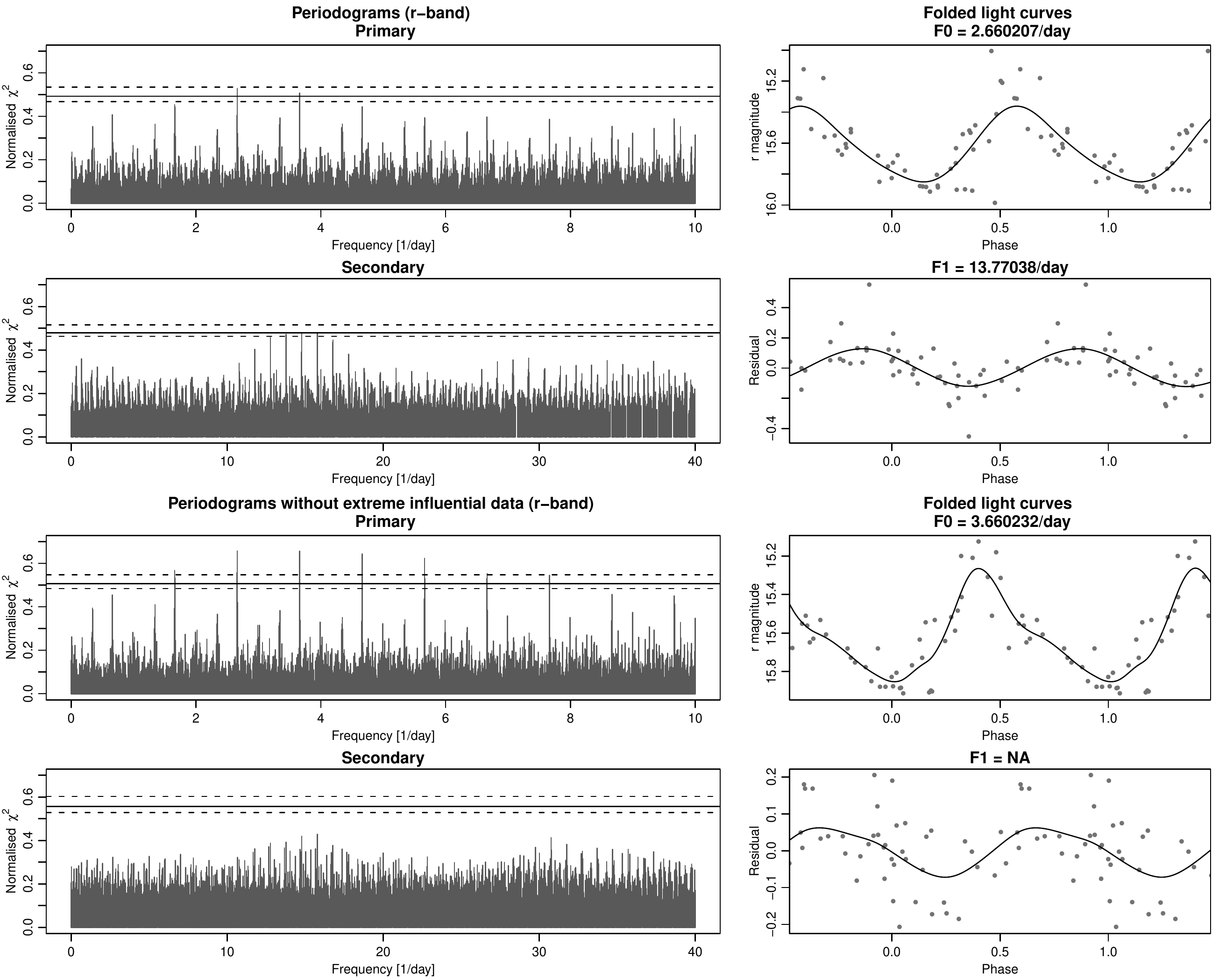}
\caption{Periodograms of $r$-band observations and of residual time series after one pre-whitening (left panels) and the corresponding folded light curves (right panels) of star 4477012 (J203120.88-001125.3). The top two rows show the periodograms and the folded light curves including the two extremal observations, the bottom rows show the same plots when omitting these from the analysis. On the periodograms, the FAP levels of 0.01 are plotted as solid black lines, together with their 95\% bootstrap confidence intervals.}
\label{fig:4477012}       
\end{center}
\end{figure*}
 
Star 4477012 (J203120.88-001125.3) also was selected as an RR Lyrae candidate in \citet{suvegesetal12b}. Its main frequency found there ($f_0 = 2.660207\; d^{-1}$), its amplitude, its principal components characteristics and the strong variation in the folded $g-i$ colour light curve suggest an RR Lyrae-like pulsating nature. The variable is on the bluest border of the RRab region of the $(u-g,g-r)$ diagram, slightly off from both RRc and RRab regions on the colour-log(period) diagram, and far from the location of both subtypes  on the period-amplitude diagram. A double-mode nature would put the star to an admissible position on the colour-log(period) plot, but this is excluded by the analysis of the residual periodogram of the $r$-band observations presented in the second row of Figure \ref{fig:4477012}: there is no indication of a correct secondary frequency. Instead, we find a weak peak at a high frequency, $f_1 = 13.77038\; d^{-1}$. It attains the level corresponding to FAP $=0.01$ only in $r$, but the maxima of the residual periodograms in all other bands falls exactly at the same frequency. This simultaneous occurrence supports that the found periodicity is a real oscillation, either of the star, or of some external origin.  

The right panel of the second row of Figure \ref{fig:4477012} shows that in the folded residual curve the largest and the smallest residual fall exactly at the minimum and the maximum of the fitted sinusoid. This is so in all bands. The found frequency is thus determined by the separation of these two extremal observations. When omitting them, the primary frequency, now at $f_0 = 3.660207\; d^{-1}$, becomes far more significant, as illustrated in the left panel in the third row of Figure \ref{fig:4477012}. From the residual periodogram, presented in the bottom row on the left, any significant peaks disappear, and the five filters show no longer marked coinciding patterns. 

\begin{figure}
\begin{center}
\includegraphics[scale=.5]{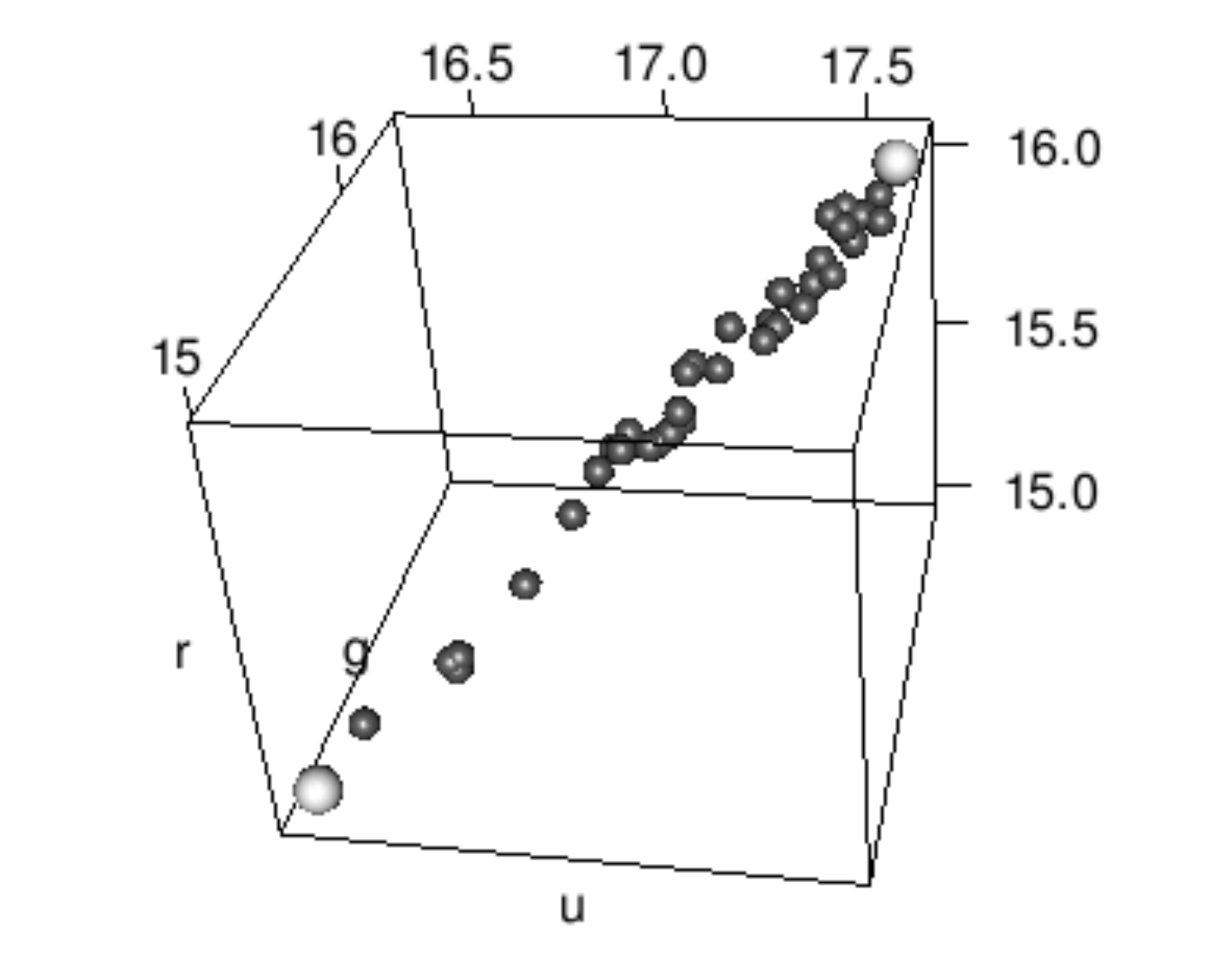}
\includegraphics[scale=.5]{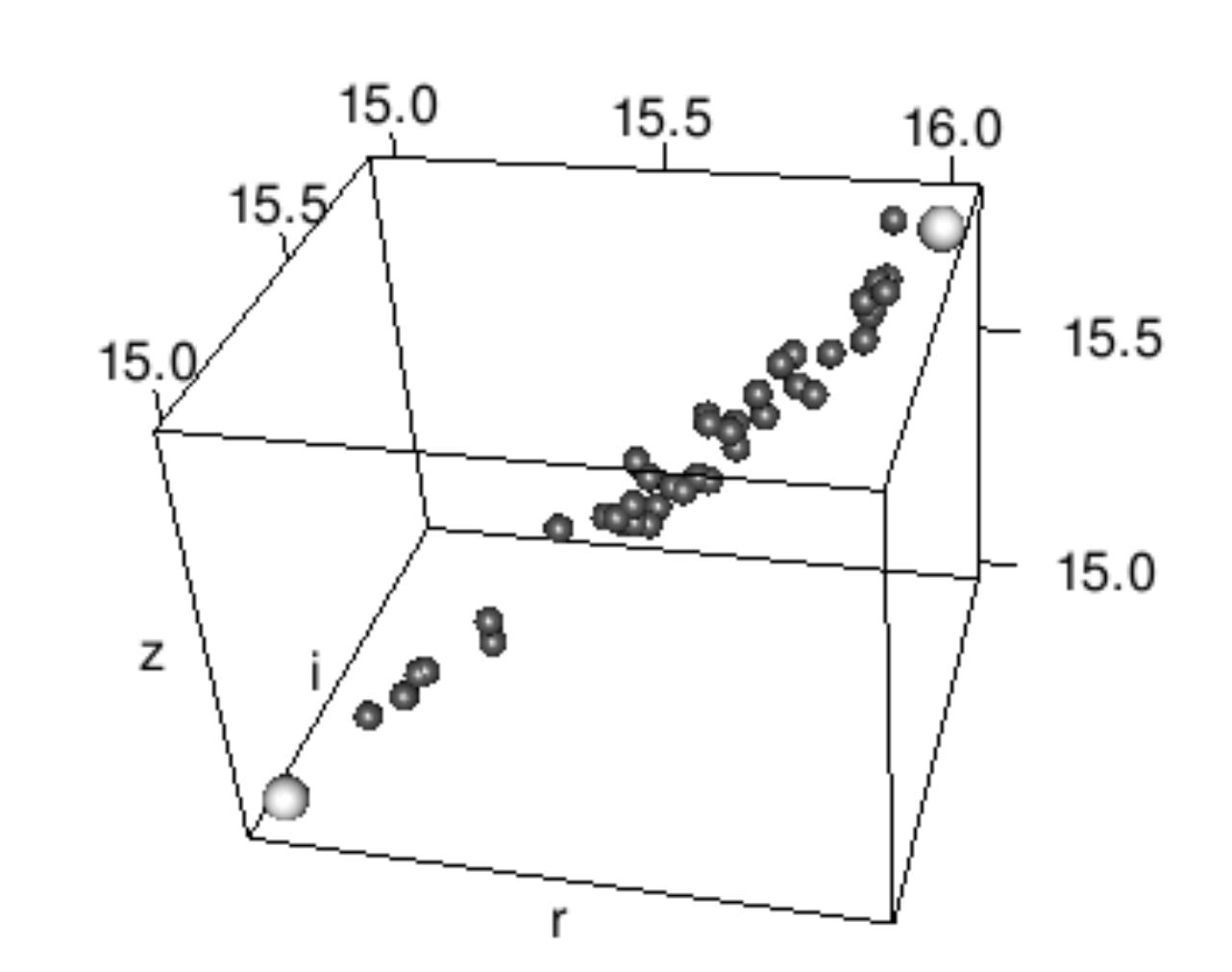}
\caption{Multivariate distribution of the 5-band observations in the $u$-, $g$- and $r$-bands (top) and in the $r$-, $i$- and $z$-band (bottom). The larger light grey blobs on both plots represent the observations with crucial influence on the secondary periodogram in Figure \ref{fig:4477012}.}
\label{fig:ugriz}       
\end{center}
\end{figure}

The decision, whether the signal of $f_1 = 13.77038\; d^{-1}$ exists or not, cannot be taken on purely statistical grounds. It depends on the decision whether the two influential observations are true or erroneous, and whether we find it plausible that only two observations out of 44 carry most of the information on an existing oscillation. These two observations are of good quality, are not outliers in any of the bands, and their error bars in all bands are rather small. Moreover, in Figure \ref{fig:ugriz}, which shows the observed ($u,g,r$) and ($r,i,z$) magnitudes of the observations in 3 dimensions plotted against each other, the light-coloured larger dots representing these points fit perfectly into the joint multivariate distribution of the remaining data. Hence, rejecting them seems unreasonable. On the other hand, accepting them and thus accepting the existence of a secondary period of such a high frequency leads to difficulties in the interpretation of the frequencies and in the class determination for the star. The confirmation or the rejection of this frequency could be obtained only by more data on this object.   

\section{Discussion}\label{sec_disc}

This paper discusses the difficulties with the most commonly used method of astronomy to estimate the False Alarm Probability in periodograms, the formula $F(z)^M$. The proposed procedure is intended to avoid its shortcomings: the lack of interpretation of $M$, the high sensitivity of the formula both to the tail of $F$ and to $M$, the invalid assumptions in its derivation, the instability of the formula due to its degeneracy when $M$ increases,  and the lack of inference for the estimated FAP or quantile levels. 

The proposed procedure combines a bootstrap of the original time series with the use of the GEV distribution instead of $F$. The bootstrap is used to produce white noise samples with similar empirical marginal behaviour as the observations, corresponding to the null hypothesis of no periodic signal in the data without imposing additional distributional assumptions. For these bootstrap repetitions, partial periodograms are computed, and the GEV is used to estimate the distribution of their maxima. The GEV, on the other hand, is a well-motivated, generally valid distribution to describe probability levels of maxima from almost all continuous  distributions. These two key elements of the procedure lift the dependence of the FAP on eventual misspecification of the periodogram distribution $F$.

The formula $F(z)^M$ is heavily impacted also by the effective number of independent frequencies, $M$. Small changes in its value change strongly the estimated quantiles or probabilities. There are no clear theoretical arguments that could help to compute its value for a given time series, or to assess whether an estimated value is reasonable or not. The proposed procedure avoids such issues. Moreover, standard asymptotic normal theory can be used to give inference about the fitted model, and to obtain confidence intervals for the parameters of the GEV and for return levels, critical levels corresponding to fixed FAP values. 

The use of the GEV also helps to relieve the computational loads due to the bootstrap, since it implies that estimates can be based on maxima of partial periodograms, computed only at a subset of frequencies. The quality of the fitted GEV model can easily be checked by diagnostic plots that are commonly used in all applications of extreme-value statistics, yielding information whether the model is good enough to use for extrapolation. This model diagnostic is imperative to do also because the astronomical periodogram is degenerate. Limiting theorems of extreme-value statistics give only sufficient conditions that do not treat this case, and though necessary conditions are likely to hold, this must be checked in all possible ways. 

The subset of frequencies at which we calculate the partial periodograms is selected in a specific way. It contains enough frequencies such that extreme-value theory, which is based on asymptotic arguments concerning maxima of a huge number of variables, can be expected to provide a good approximation. Intervals of contiguous frequencies are randomly selected across the tested frequency range $(0,f_{\max}]$, each of length at least equal to the oversampling factor. This ensures that they sample the most important sorts of dependencies in the periodogram: the long-range correlations across the whole frequency range, and the extremely strong local dependence due to spectral leakage.

The results on simulations show that the provided FAP levels are reasonable, and convey reliable information on the plausibility of the null hypothesis. The confidence intervals on the critical levels provide further useful information on the uncertainty of the decision. The estimated quantile levels are remarkably stable both with respect to the number of bootstrap repetitions and the size of the frequency subset, though a somewhat higher number of bootstrap repetitions provides better results for short ($N=25$) time series, and larger numbers produce of course less uncertainty of the estimates. A setup that provides good estimates and reasonably narrow confidence intervals on the quantile levels, but computationally is not too heavy, takes only a few times the time of the calculation of the complete periodogram (less than 10 times both in the simulations and the data examples). The procedure was also applied to two variable stars with unknown class from SDSS Stripe 82, which had marginally significant primary periods in the RR Lyrae range. One of these is proved to be a double-mode RR Lyrae, by the emergence of a highly significant secondary peak at a frequency that corresponds to the Petersen diagram of the double-mode RR Lyrae stars. The other star remains an unclear case. A barely significant secondary period was found at a high frequency. The weak significance is due to the fact that this period is mainly determined by only two observations. However, these observations are very unlikely to be erroneous measurements, so the presence or absence of this high-frequency variation in this star cannot be decided purely on statistical grounds.

The new procedure, because of its stability, well-founded statistical background and broad applicability, is therefore an excellent alternative to obtain significance of periodicities detected in astronomical periodograms.

\section*{Acknowledgments}

The author thanks R. Ga\'al for interesting discussions, and P. Dubath and L. Rimoldini for many valuable comments on the manuscript. The work was partly supported by the Swiss National Science Foundation grant  PMPDP2\_129178.


\bsp

\label{lastpage}

\end{document}